\documentclass[a4paper,11pt]{article}
\pdfoutput=1 % if your are submitting a pdflatex (i.e. if you have
\usepackage{jinstpub} % for details on the use of the package, please
\usepackage{lineno}
\usepackage{amsthm}
\usepackage{mathtools}
\usepackage{graphicx}
\usepackage[utf8]{inputenc}
\usepackage{amsmath}
\usepackage{caption}
\usepackage{subcaption}

\graphicspath{{used_figures/}}

\title{\boldmath Simulation of RPC responses in mini-Iron Calorimeter for cosmic ray muon studies}

\author[a,b,1]{J.M. John,\note{Corresponding author.}}
\author[b]{G. Majumder,}
\author[c]{S. Pethuraj}

\affiliation[a]{Homi Bhabha National Institute,\\Mumbai 400094, India}
\affiliation[b]{Tata Institute of Fundamental Research,\\Mumbai 400005, India}
\affiliation[c]{PSNA College of Engineering and Technology,\\Dindigul 624622, India}

\emailAdd{jim.john@tifr.res.in}

\abstract{A Prototype of the Iron Calorimeter for the India-based Neutrino Observatory is currently running in Madurai, India. This consists of twenty large area single gap Resistive Plate Chambers (RPCs) of size,  $\sim$ 185\,cm $\times$ 175\,cm sandwiched between 11 layers of iron plates of thickness 5.6\,cm. The total size of the mini-Iron Calorimeter(mini-ICAL) is 4\,m$\times$4\,m$\times$1.2\,m and having a weight of 85\,tons is a (1/600) scaled-down version of the ICAL detector of INO~\cite{ichep2018}. The detector is magnetized using iron plates and two sets of copper coils, each coil has 18 turns. The central region of the mini-ICAL, where the RPCs are placed, has a nearly uniform magnetic field of 1.4\,T, similar to the magnetic field in the final ICAL detector. This prototype serves as the basis for the design of the demonstrators, which closely mimics the characteristics of a future ICAL. This detector is built to test the final electronics in the fringe field of the magnet and also to develop the experience to construct the ICAL detector.

The mini-ICAL has been operational since 2018 and collects cosmic muon data with different configurations of RPC, and electronics, which are continuously changing for various R\&D efforts. Dedicated efforts are made in parallel to  measure the momentum and azimuthal angle  of $\mu^+$ and $\mu^-$ independently, which could be an important input to the cosmic neutrino event generators. To improve the precision of those measurements a dedicated simulation effort was made, particularly in the digitisation of the RPC signal for a real detector simulation, where efficiency, noise rate, strip multiplicity, etc. were matched with the real data collected during runs at mini-ICAL.}


\begin{document}
\maketitle
\flushbottom

\section{Introduction : An Overview of the Iron Calorimeter Prototype}
\begin{figure}[htbp]
\centering % \begin{center}/\end{center} takes some additional vertical space
\includegraphics[width=1.0\textwidth]{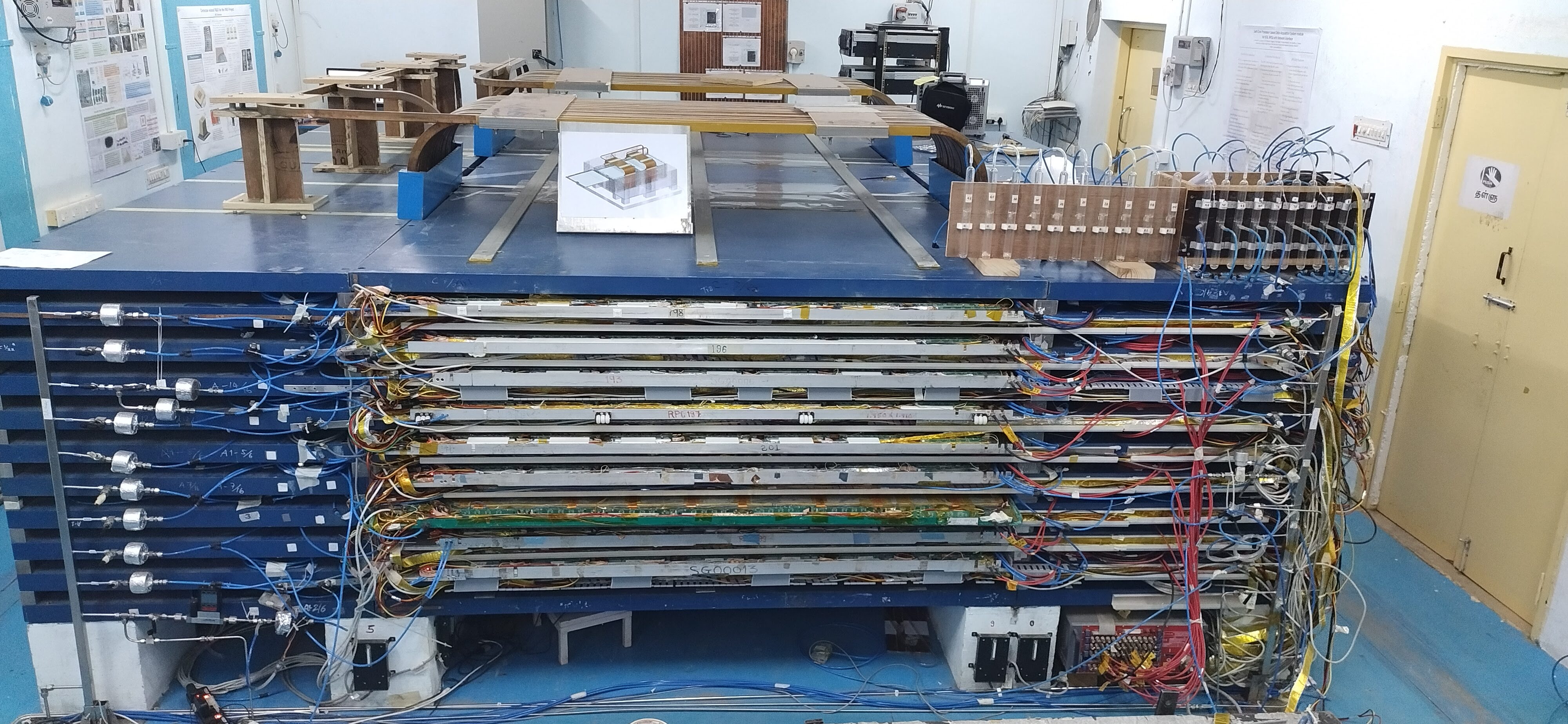}
\caption{\label{fig:miniICAL_original} The fully assembled mini-ICAL Detector.}
\end{figure}

\begin{figure}[htbp]
\centering % \begin{center}/\end{center} takes some additional vertical space
\includegraphics[width=.7\textwidth]{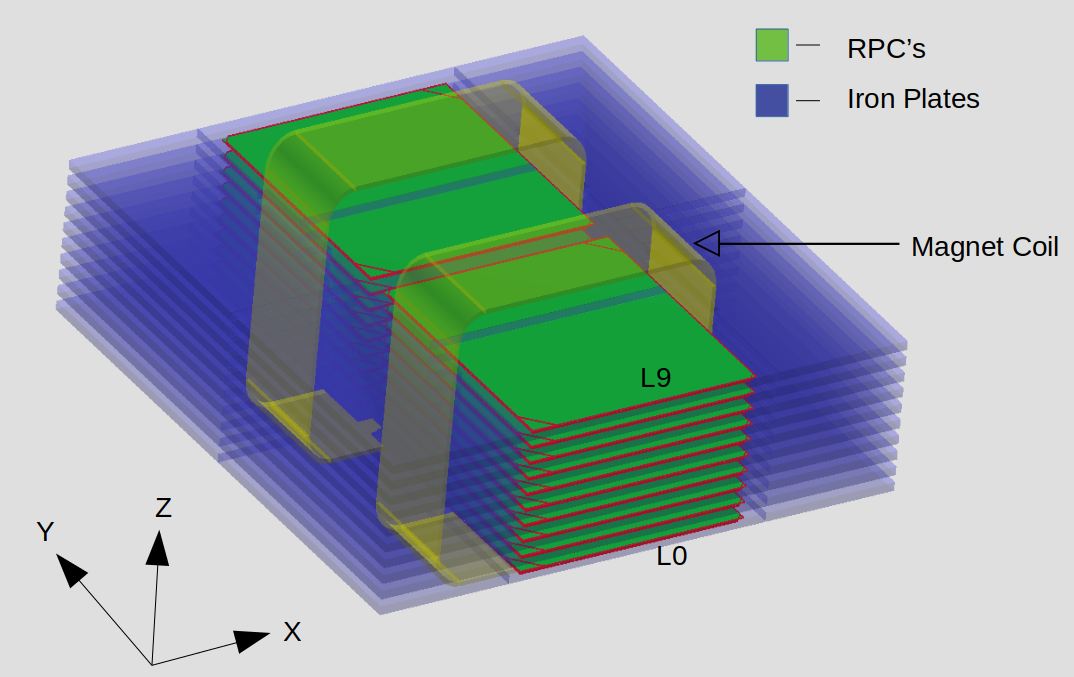}
\caption{\label{fig:miniICAL} The RPC positioning inside mini-ICAL in Geant4 detector geometry(Adapted from \citep{jim_paper1}).}
\end{figure}

The prototype detector of the ICAL i.e., mini-ICAL is currently in operation at IICHEP,
Madurai~\cite{ichep2018} and shown in Figure~\ref{fig:miniICAL_original}. The mini-ICAL detector is a (1/600)th scaled-down version of the ICAL and consists of twenty 2\,m $\times$ 2\,m glass RPCs sandwiched between 11 layers of iron plates as shown in Figure~\ref{fig:miniICAL}. The detector is magnetized using iron plates and two sets of copper coils, each coil has 18 turns. The central region of the mini-ICAL, where the RPCs are placed, has a nearly uniform magnetic field of 1.4\,T for a coil current of 900 amps. More details about the mini-ICAL detector can be found at \cite{ichep2018}. The mini-ICAL detector mainly tracks the cosmic muons and identifies their four vectors with the help of the magnetic field. The purpose of building mini-ICAL was to gain experience in RPC handling and operation, operating gas system, operation of magnet and its cooling system, and to check the performance of electronics (trigger system, DAQ, and front end) in the presence of the magnetic field.

 \section{RPC in mini-ICAL}
Each of the iron plates in mini-ICAL is 5.6\,cm thick. The industrial production of iron plates had a flatness as specified (within the specification of $\pm$3\,mm over 2\,m $\times$ 2\,m). Hence, the iron plates are stacked one above another with a gap of 4.5\,cm though the required space for the RPC was less than 4\,cm and the RPCs are placed in the central region (2\,m$\times$4\,m) of the mini-ICAL. RPCs are made by placing two thin glass plates of size 185\,cm $\times$ 175\,cm and 3\,mm thickness, placed 2\,mm apart from each other. The bulk resistivity of the used glass is 3 - 4$\times$10$^{12}\, \Omega$cm~\cite{raveendrababu}. The gap between the two glass plates is maintained to be 2\,mm using circular polycarbonate “button” spacers of diameter 11\,mm. The gas mixture used for RPC operation is R134a (95.2\%), Isobutane (4.5\%), and SF$_6$ (0.3\%). There are two sets of inlet and outlet nozzles through which the gas mixture is flown with a rate of 6\,SCCM. The outer surfaces of the glass chamber are coated with a thin film of graphite paint. Graphite paint is applied by the screen-printing technique. The surface resistivity of the graphite coat is in the range of 0.6-1.5\,M$\Omega/\square$. This graphite coating permits the application of a high voltage across the chamber as well as allows induced electrical signals to pass through it. The RPCs are operated at 9.8\,kV to 10.2\,kV based on the efficiency plateau using the cosmic muon data~\cite{achrekar}. Two thin layers (100 $\mu$m) of Mylar sheets (insulator) are kept in between the chamber and pick-up panels on both sides. The pick-up panel is made of copper (readout) strips with a width of 2.8\,cm and an inter-strip gap of 0.2\,cm pasted on a plastic honeycomb structure. The other side of the pick-up panel is pasted with thin aluminum foil and is connected to the ground. This acts as a Faraday cage for the strips. The readout strips are placed orthogonally on either side of the RPC to locate the position of the traversed particle. There are a total of 58 strips on the X-Side (bottom pickup panel) and 61 strips on the Y-Side (top pickup panel). The upwards direction is taken as the Z-coordinate of the mini-ICAL, the Y-coordinate is taken towards the back side wall as shown in Figure~\ref{fig:miniICAL} and the X-coordinate is the right side, considering the left-handed coordinate system. The magnetic field is along the Y-axis, thus in the presence of a magnetic field, muon trajectories bend in the XZ plane. RPC layers are numbered from bottom to top, i.e., the bottommost layer is assigned as Layer-0 and the topmost as Layer-9 as shown in Figure~\ref{fig:miniICAL}.
 
  The induced signals are amplified and discriminated by an 8-channel NINO ASIC chip~\cite{nino}. The low-voltage differential signaling output of NINO is fed to the FPGA-based RPC data acquisition system. The Digital Front End module consists of several functional blocks such as a Time-to-Digital Converter (TDC), Strip-hit latch, Pre-trigger generator, Rate Monitor, Front End control, and ambient parameter monitor. A pre-trigger signal was generated by the OR of all strips on either the X-side or Y-side and the 'OR' of the X-side and Y-side trigger is used to store the data. A second-level trigger for cosmic muons was created independently for the X-side and Y-side by requiring the coincidence of pre-trigger signals in layers 6, 7, 8, and 9 (top four layers of RPC) within a time window of 100 ns. A typical trigger rate of this setup is $\sim$180\,Hz. Once the trigger is formed the digitized data is transmitted to the back-end using the DAQ boards network interface~\cite{mandar}. More detail on this setup is available in~\cite{jim_paper1}. For the last four decades, RPCs have been used in high-energy physics experiments. But, in those experiments, the single gap RPCs were primarily used for triggering and the multigap RPCs were used for the timing. This is the first attempt to measure the momentum using an experimental setup with only RPC information. The strip multiplicity due to the muon signal as well as noise in RPCs largely depends on the fabrication of the RPC, the electronics, as well as running conditions, e.g.,  applied HV, temperature, humidity, the composition of gas mixture as well the presence of contaminations, etc. Thus, it is essential to have a detailed simulation of RPCs based on the experimental condition, which is not available in the literature.

\section{Properties of the RPC detector during the operational phase}
\begin{figure}[htbp]
     \centering
     \begin{subfigure}[b]{0.3\textwidth}
         \centering
         \includegraphics[width=\textwidth]{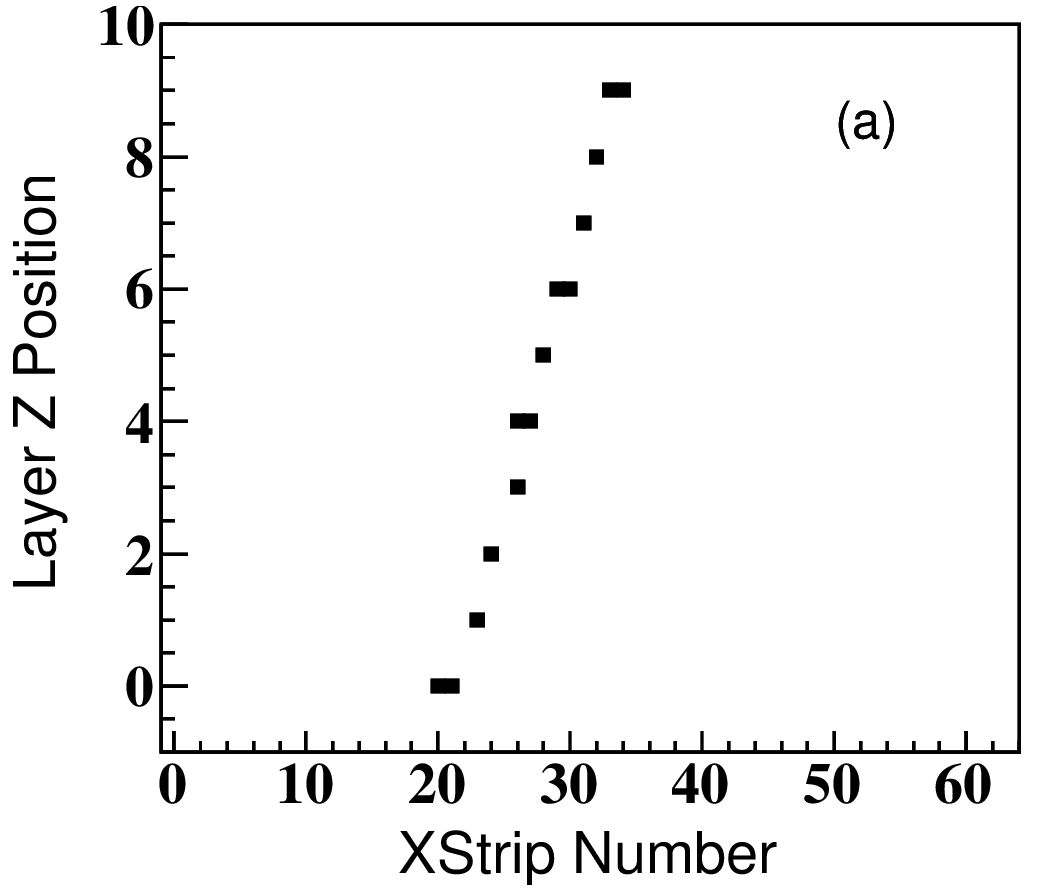}
     \end{subfigure}
     \hfill
     \begin{subfigure}[b]{0.3\textwidth}
         \centering
         \includegraphics[width=\textwidth]{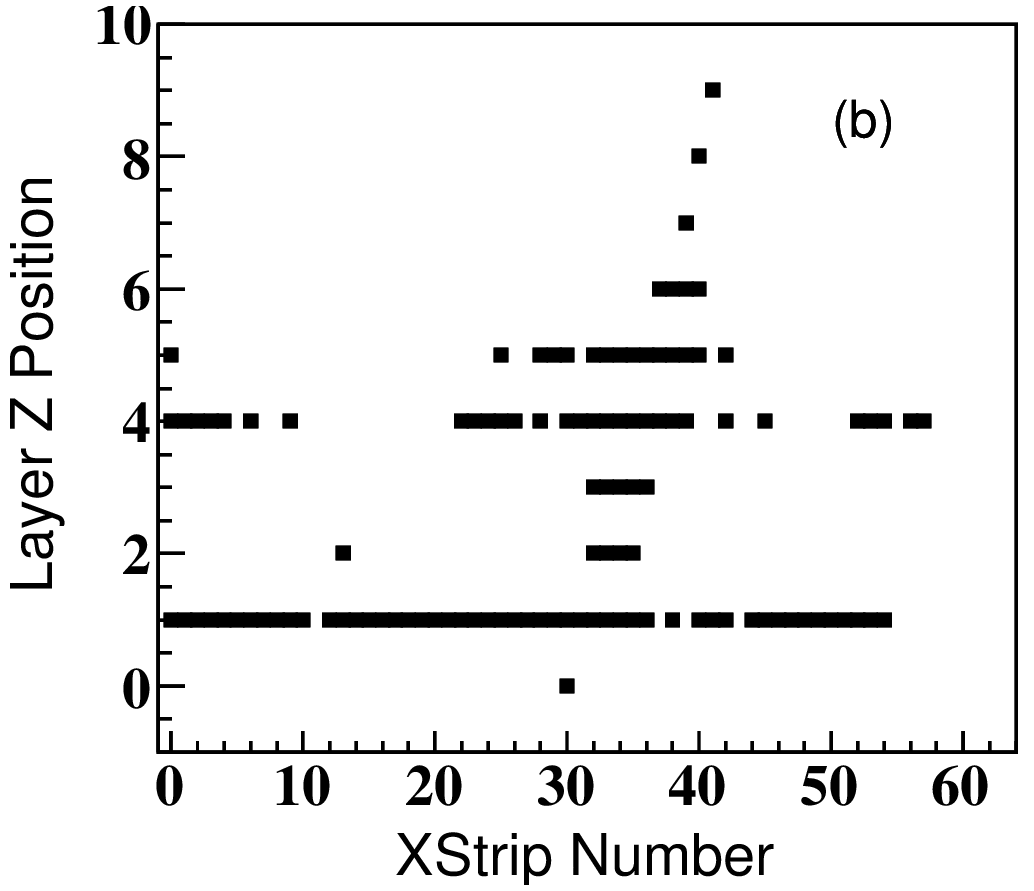}
     \end{subfigure}
     \hfill
     \begin{subfigure}[b]{0.3\textwidth}
         \centering
         \includegraphics[width=\textwidth]{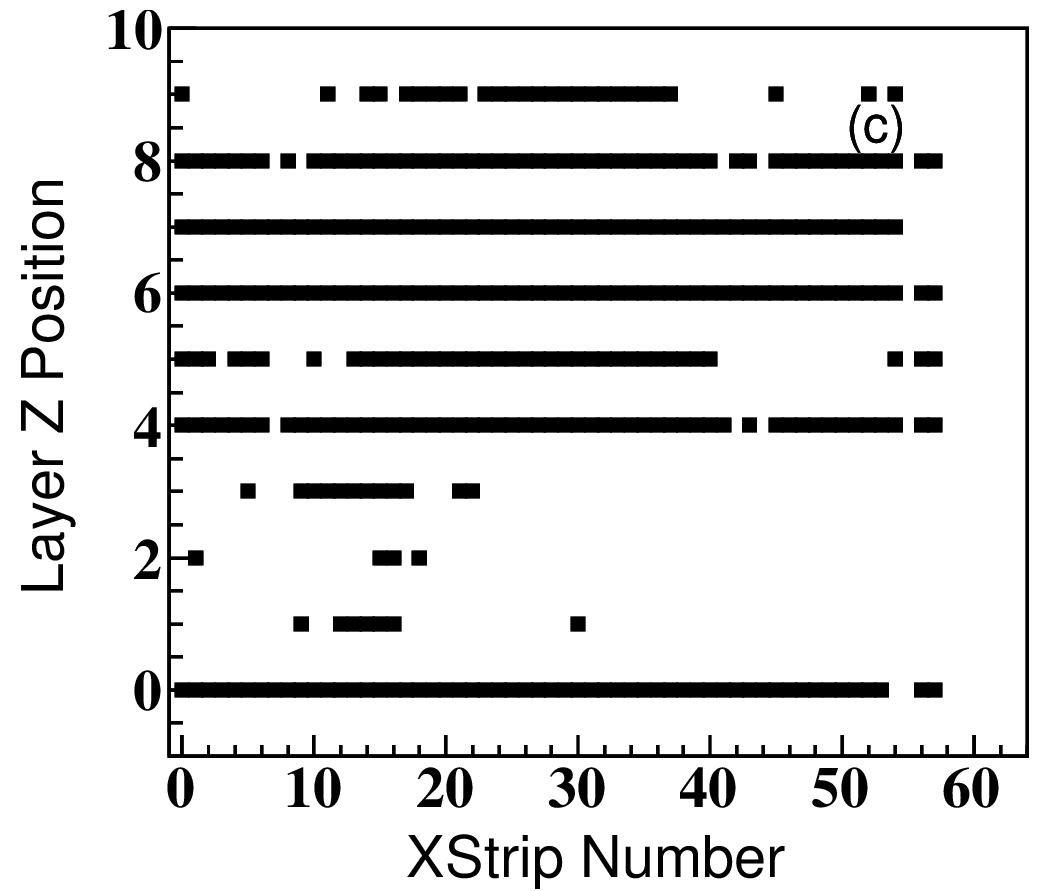}
     \end{subfigure}
        \caption{Figure shows three different categories of events in mini-ICAL, (a) Simple muon hit (b) Hadron shower (c) Event triggered by noise (Trigger layers are 6,7,8, and 9).}
        \label{fig:three_events}
\end{figure}

The data used for the real detector simulation were taken in December 2018 with and without magnetic fields. In general, during the day time, the magnet was ON, and during night time data was taken without a magnetic field. There are mainly three kinds of events in the triggered data, i.e., $(i)$ event with only one muon (or pion) trajectory (occasionally multiple muon trajectory,~\cite{surya_multiplicity}), $(ii)$ hadronic showers, which are mainly due to shower development in the roof and/or in the mini-ICAL and $(iii)$ trigger generated due to the electronic noise. Typical events are shown in Figure~\ref{fig:three_events}. This study is based only on the first category, e.g., a single muon event. Though the algorithm is able to manage multiple trajectories in an event, a restriction was imposed to study the properties of single muon event. As is seen in Figure~\ref{fig:three_events}(a) there are 2-3 hits \footnote{Induced signal in the strip is typically of the order of few hundred fC in RPCs, signal above 100\,fC are considered as a hit or a measurement. The average noise pulses measure around 10\,fC.} due to muon trajectory, which is common in the RPC detector. For a good position resolution, a hit multiplicity of more than three is avoided \cite{jim_paper1}. In the presence of noise in a layer, there are possibilities that other hits might be visible in some other strips far away from the muon trajectory. Those layers are used for the study of the performance of an RPC, but not included for the measurement of the four vectors of muon trajectory.

  \subsection{Efficiency, multiplicity and resolution of RPC detector}
\begin{figure}[htbp]
     \centering
     \includegraphics[width=\textwidth]{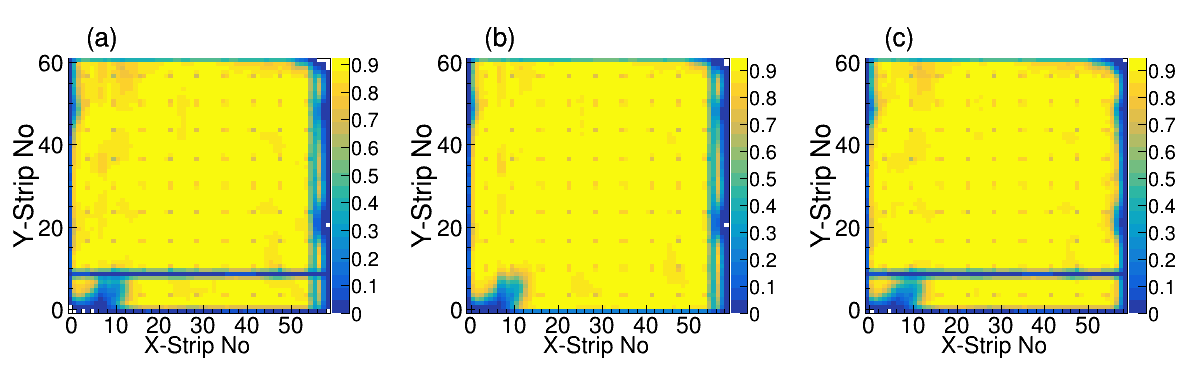}
        \caption{Figure shows efficiency of Layer 5, (a) Correlated efficiency (b) X-Side efficiency (c) Y-Side efficiency.}
        \label{fig:efficiency}
\end{figure}

  The Performance of the RPCs was studied in data both with and without magnetic field and it was found that the noise rate and efficiency remain the same in both cases. Thus, for simplicity, initially, most of the performances and comparisons were done with data without any magnetic field. All the properties are different for different RPCs in the stack, hence they have to be extracted and reincorporated independently. The selected clusters (combinations of one to three consecutive hits in a layer) from different layers were fitted with a straight line in XZ and YZ views independently. The chamber under study was not included in that fit. Using the deviation of the extrapolated position of the muon trajectory and the measured position of the signal, an iterative method was used to align all RPCs~\cite{Pethuraj_2017}. The extrapolated/interpolated muon position in the chamber is considered as the muon position and the position-dependent efficiency of the detector is estimated using the presence or absence of hits near that position. Figure~\ref{fig:efficiency} illustrates a typical example of efficiency observed in layer 5. In Figure~\ref{fig:efficiency}(a), the efficiency is depicted when there are signals present on both the X-side and Y-side. Additionally, Figure~\ref{fig:efficiency}(b) showcases the efficiency with a signal on the X-side, while Figure~\ref{fig:efficiency}(c) demonstrates the efficiency with a signal on the Y-side independently. The typical uncorrelated efficiency varies from 82--88(79--82)~$\%$ for the X(Y) side and the correlated efficiency varies from 79--82~$\%$ in our stack. The inefficiency in the bottom left corner of the RPC is probably due to island formation in the graphite coating, which is due to painting issues and the high voltage is not applied in that region. The 9$^{th}$ strip in the Y-side is dead due to the malfunctioning of the electronics chain.
  
\begin{figure}[htbp]
     \centering
     \begin{subfigure}[b]{0.45\textwidth}
         \centering
         \includegraphics[width=\textwidth]{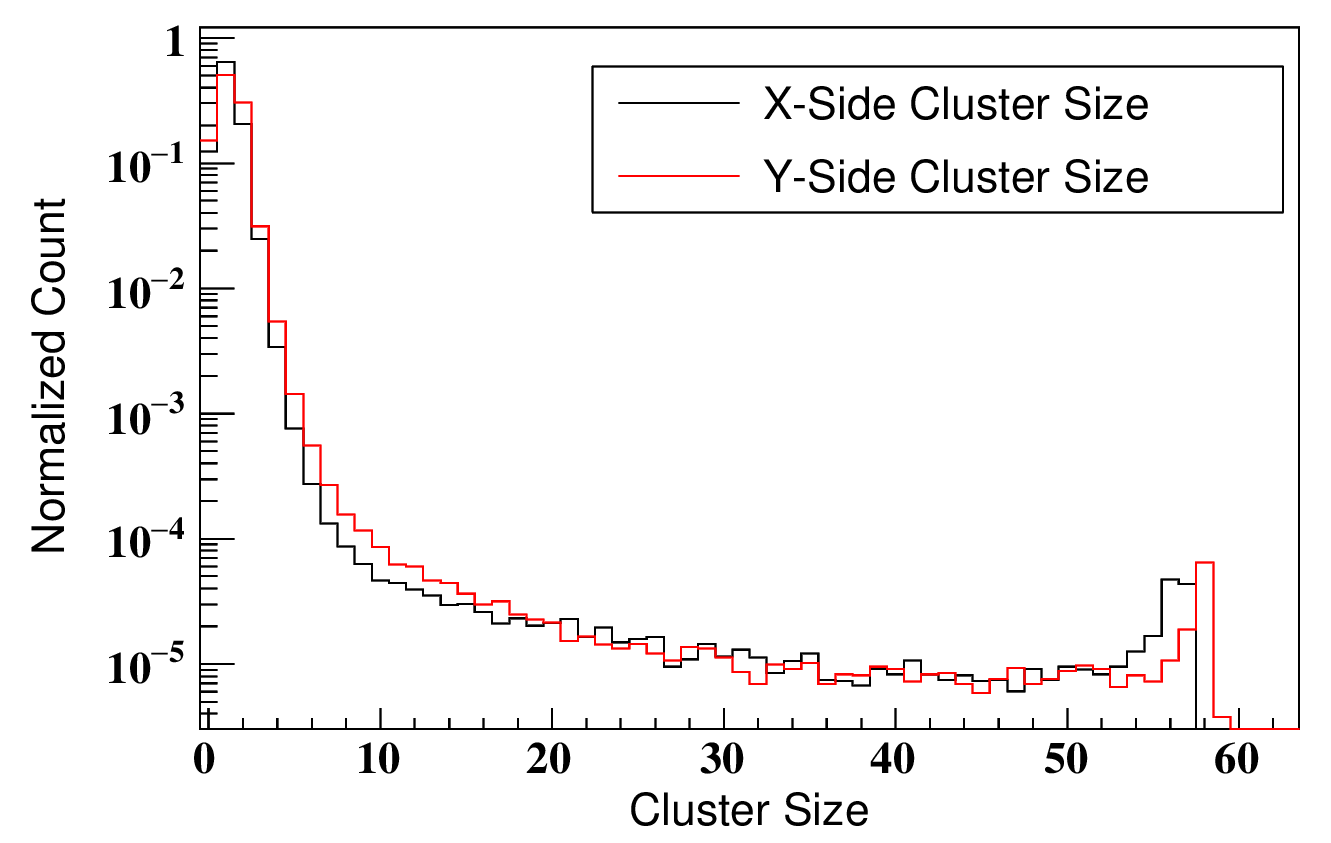}
     \end{subfigure}
     \hfill
     \begin{subfigure}[b]{0.49\textwidth}
         \centering
         \includegraphics[width=\textwidth]{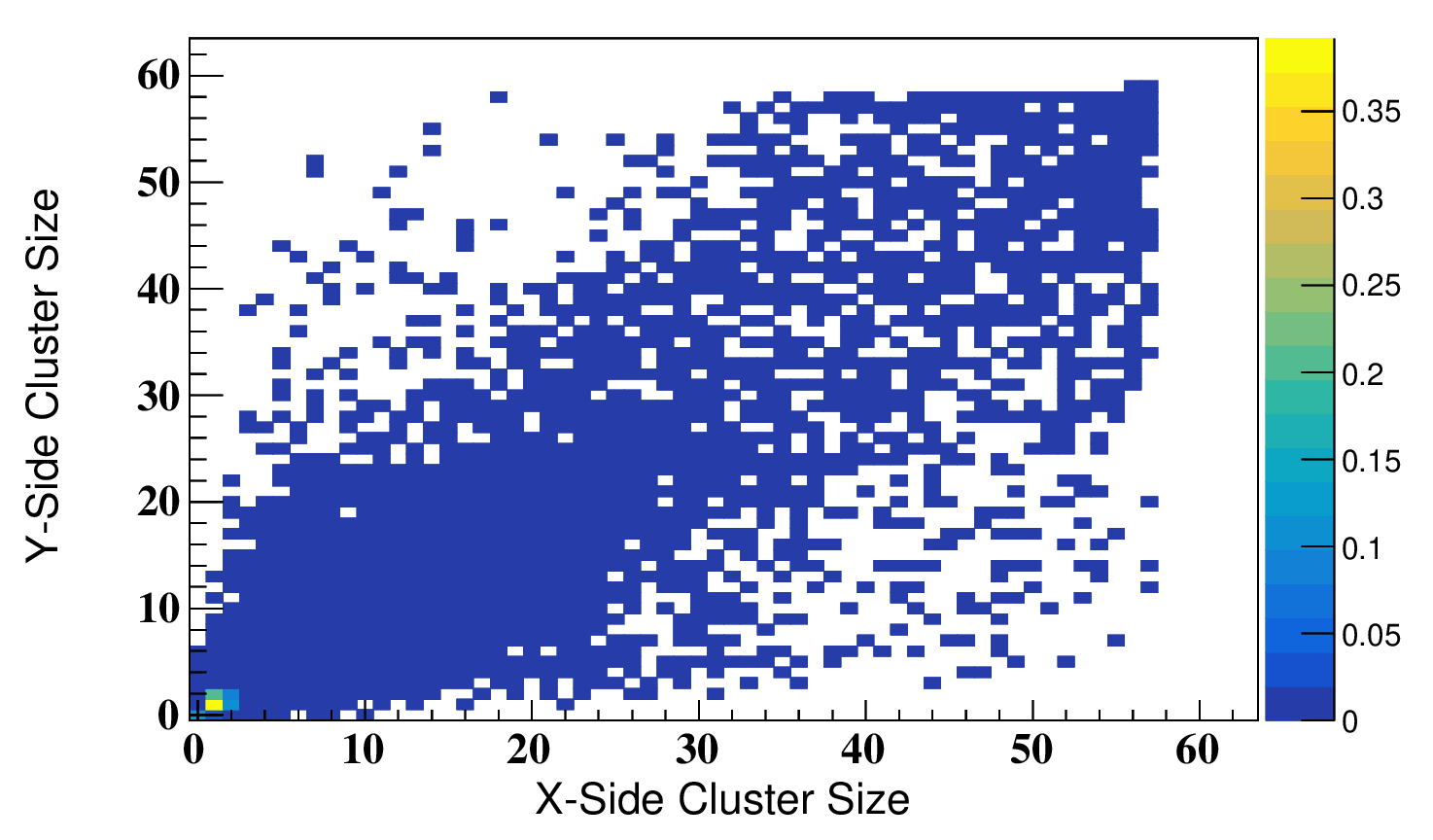}
     \end{subfigure}
     \caption{Figure shows cluster size of X-Side and Y-Side in Layer 5 and its correlation.}
        \label{fig:multiplicity}
\end{figure}

The overall multiplicities in X- and Y-sides are shown in Figure~\ref{fig:multiplicity} along with the correlations of multiplicities. The long tails of large strip multiplicities, more than 20 are mainly due to correlated electronics noise, while multiplicities below three are primarily due to avalanche formation inside the RPC chamber. Intermediate multiplicities are due to streamer formation and/or electronic noises, where one loses the localisation of the signal.

\section{Simulation}
The simulation framework uses a combination of the CORSIKA (COsmic Ray SImulations for KAscade) \cite{corsika} and GEANT4 toolkits \cite{geant4}. The CORSIKA simulation has been employed to model the extensive air shower with the magnetic rigidity cutoff for the location of the study. The hadron interaction models FLUKA and SIBYLL are respectively used in the lower and higher energy range. The primaries are generated from 10\,GeV to 10$^6$\,GeV for different primaries (H, He, C, O, Si, and Fe) with a spectral index of $-$2.7. The zenith and azimuth angle of primary particles are generated uniformly within the range of 0 - 85$^{\circ}$ and 0 - 360$^{\circ}$ respectively but are allowed to propagate depending on the rigidity cutoff in different ($\theta$,$\phi$) bins at the location of the mini-ICAL detector. The particles generated by the CORSIKA at the observation surface are provided as input to the detector simulation. The observation plane has been divided into squares of the size of 4\,m$\times$4\,m and an event includes all particles passing through the square. This paper presents the MC simulation where the state of the detector is extracted and reincorporated back into the simulation, e.g., noise associated with muons is extracted from the data and incorporated in the MC, which is presented in  Section~\ref{section:noise}, where the state of all the 10 RPCs are extracted independently and reincorporated.

\section{Response to muons and Cluster size simulation}
\label{section:multiplicityandeffi}
The efficiency for particle detection in RPCs is not uniform, with the center of the strips exhibiting higher efficiency compared to the edges. There is increased inefficiency towards the edges of the RPCs and areas where spacers are located. The response of each electronic component varies, and in the case of RPCs, it typically consists of a correlated component on both the X and Y sides, as well as an uncorrelated component. The correlated component in RPCs arises from the internal processes within the RPC itself, while the uncorrelated component stems from the events occurring after the signals have been induced in the pickup panels. So the response also has to be simulated in the same way.

\begin{figure}[htbp]
\centering % \begin{center}/\end{center} takes some additional vertical space
\includegraphics[width=.9\textwidth,origin=c]{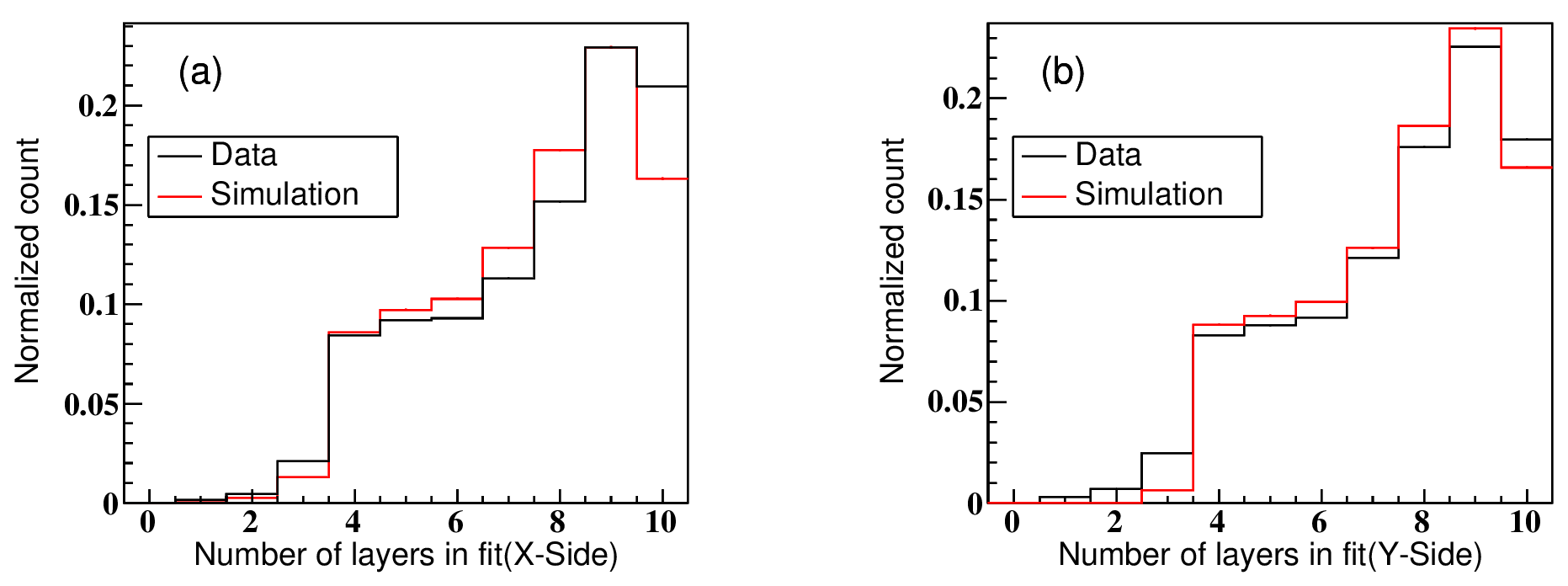}
% "\includegraphics" from the "graphicx" permits to crop (trim+clip)
% and rotate (angle) and image (and much more)
\caption{\label{fig:simulation_gen_ndf} Comparison of number of layers in muon trajectories between Data and MC where MC is simulated using the efficiency map as shown in Figure~\ref{fig:efficiency}(a) and cluster size distribution is independent of global co-ordinate in RPC, but depends on the local co-ordinates in the strips (a) X-side, (b)  Y-side.}
\end{figure}

The accuracy of position resolution is greatly influenced by the distribution of cluster sizes. To ensure precise matching of position residues, it is imperative to re-establish the relationship between cluster size and position within a strip. The correlation between cluster sizes and the position of the muon passing through a strip is evident. When the muon traverses through the center of the strip, there is a significantly higher likelihood of having a cluster size of 1 compared to a cluster size of 2. An effort was made to replicate this phenomenon in simulation using the efficiency map of all the RPC's and cluster sizes are determined by the position in the strip. However, this attempt proved unsuccessful in reproducing the number of layers with valid hits of a reconstructed muon trajectory, as depicted in Figure~\ref{fig:simulation_gen_ndf}. Here the noise distribution specified in section~\ref{section:noise} is included in the simulation.

\begin{figure}[htbp]
\centering % \begin{center}/\end{center} takes some additional vertical space
\includegraphics[width=.9\textwidth,origin=c]{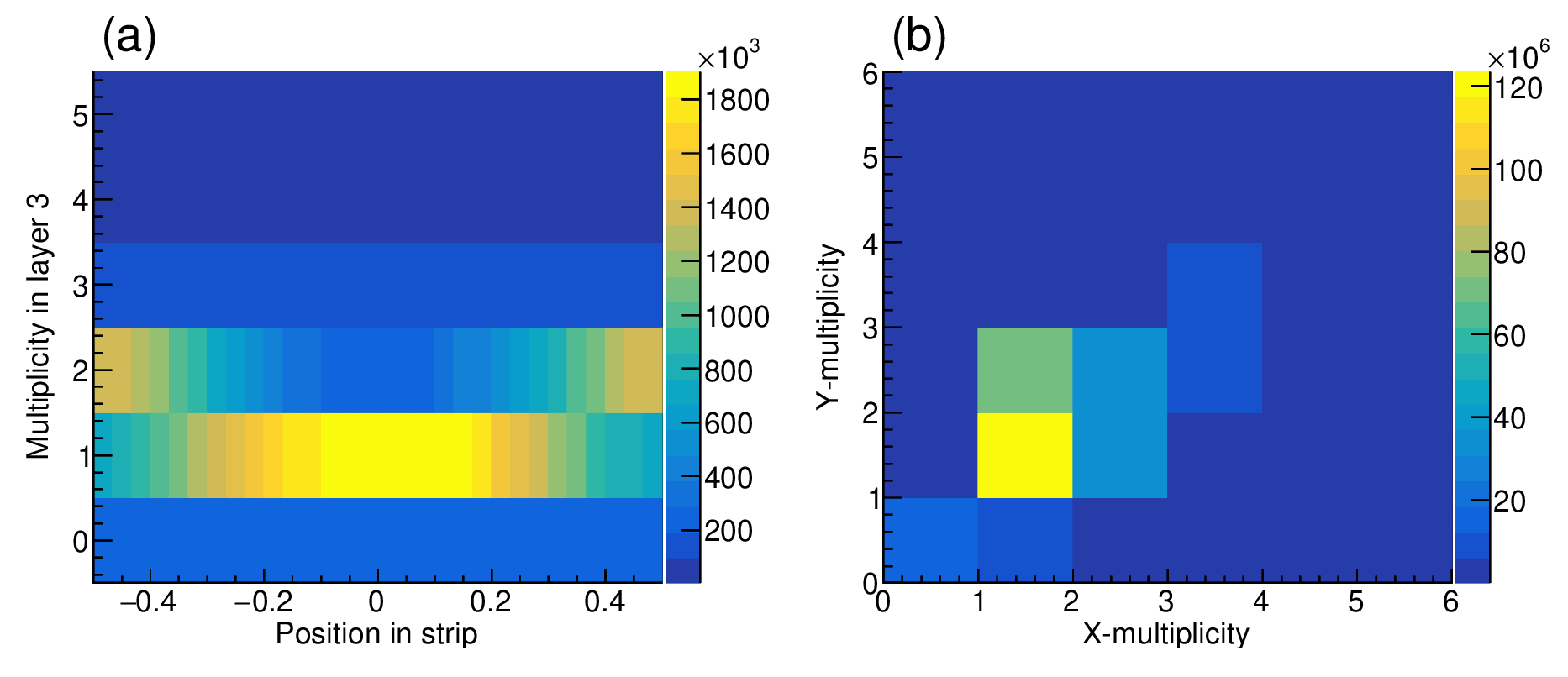}
% "\includegraphics" from the "graphicx" permits to crop (trim+clip)
% and rotate (angle) and image (and much more)
\caption{\label{fig:clustersize} (a) The Cluster size with respect to position in a strip,
  (b)The Joint distribution of X-side cluster size and Y-side Cluster size.}
\end{figure}

In order to replicate the cluster sizes and efficiencies accurately, each RPC is divided into 64 $\times$ 64 blocks, and information regarding the cluster size corresponding to the position within a strip is recorded in a two-dimensional histogram. This distribution is illustrated in Figure~\ref{fig:clustersize}(a), facilitating a comprehensive analysis of the cluster size distribution and its relationship with position.
 
The aforementioned two-dimensional histogram is created independently for both the X-side and Y-side of the RPC. As the X-side and Y-side exhibit correlation due to internal processes within the RPC, it is crucial to reintroduce this correlation. In this situation, both uncorrelated and correlated inefficiency have been recreated using the first bin of the x-side and y-side in the two-dimensional joint distribution depicted in Figure~\ref{fig:clustersize}(b).

The relationship between the X-Multiplicity and Y-Multiplicity appears to exhibit a dependency pattern reminiscent of a Clayton copula, as depicted in Figure~\ref{fig:multiplicity}. Equation~\ref{eqn:clayton} defines the Clayton copula, where 'u' and 'v' denote two uniform random numbers. 
\begin{equation}
C(u,v) = \Big[u^{-\alpha} +v^{-\alpha}-1\Big]^{-1/\alpha}
\label{eqn:clayton}
\end{equation}

\begin{figure}[htbp]
\centering % \begin{center}/\end{center} takes some additional vertical space
	\begin{subfigure}[b]{0.3\textwidth}
        \centering
    	\includegraphics[width=\textwidth]{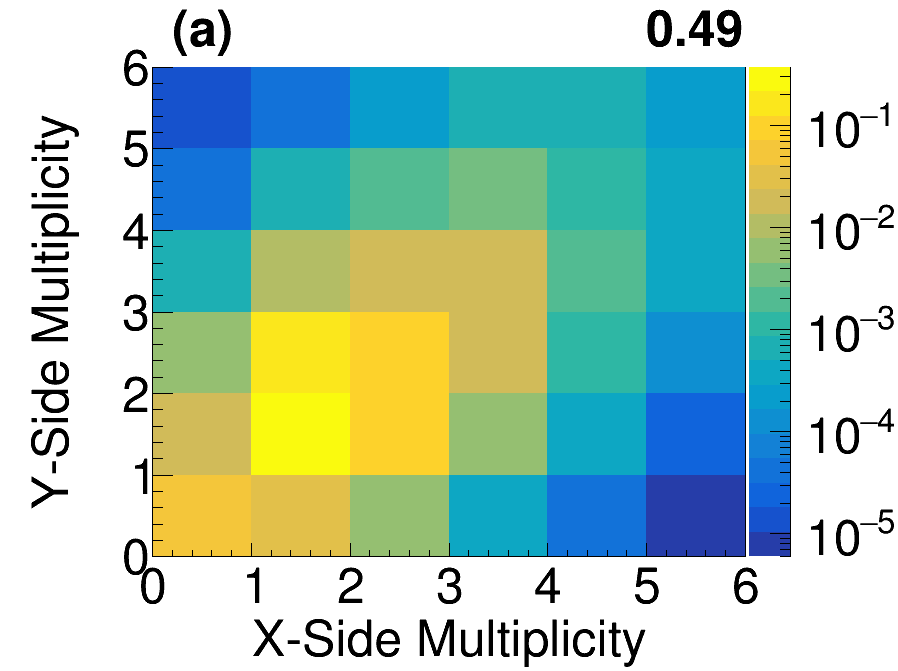}
    \end{subfigure}
\hfill
	\begin{subfigure}[b]{0.3\textwidth}
         \centering
         \includegraphics[width=\textwidth]{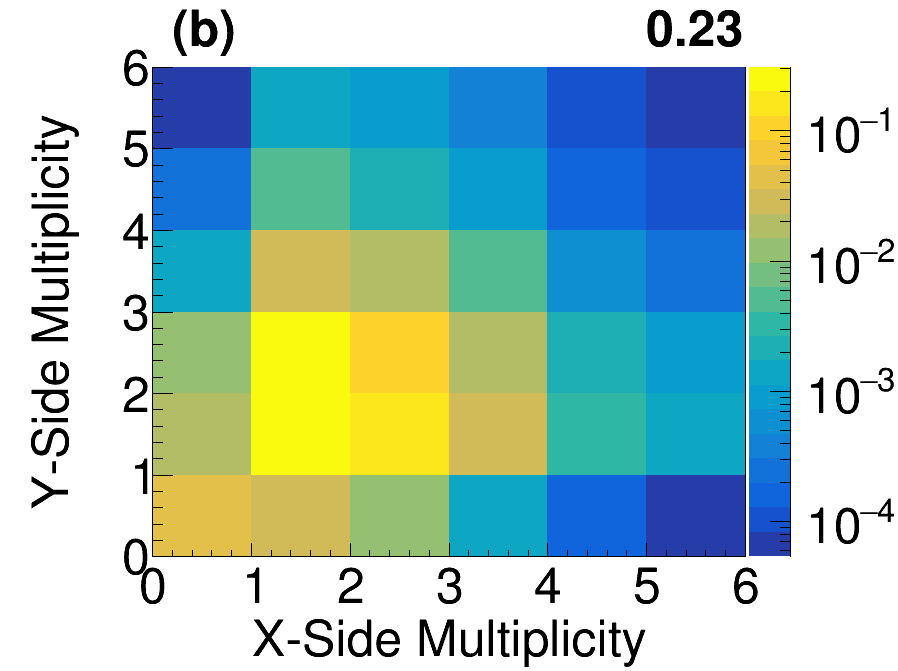}
    \end{subfigure}
\hfill
	\begin{subfigure}[b]{0.3\textwidth}
         \centering
         \includegraphics[width=\textwidth]{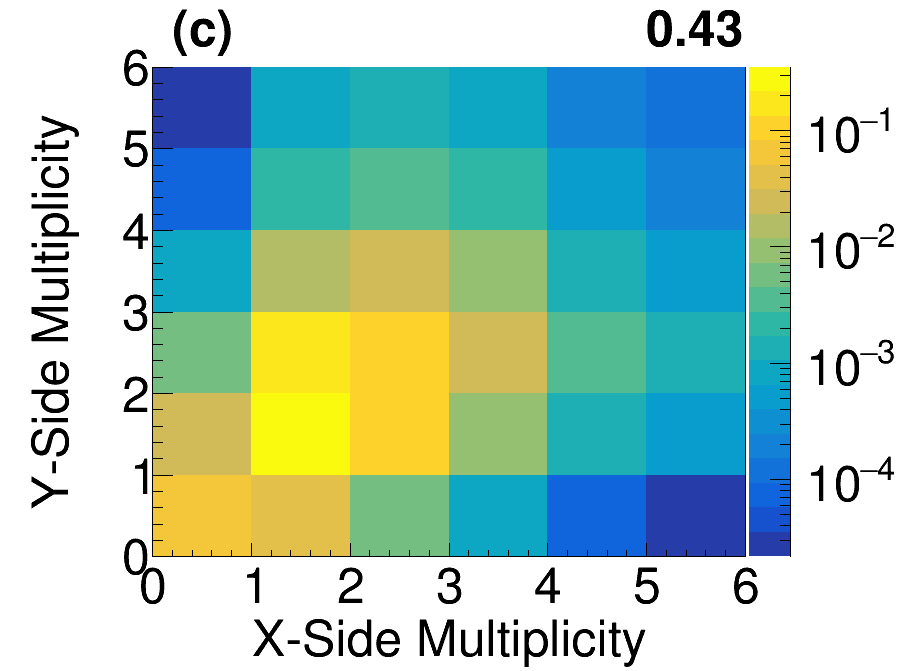}
    \end{subfigure}
\caption{\label{fig:joint_dist} The joint distribution of X-side cluster size and Y-side cluster size is illustrated in three scenarios: (a) based on data, (b) from simulation without correlation, (c) from simulation, after incorporating correlation using a Clayton copula ($\alpha$ = 3) for dependence. The correlation factor is displayed in the top right corner of every figure as well.}
\end{figure}

By solving this equation, two correlated uniform random numbers are generated. By adjusting the parameter $\alpha$, the joint distribution of multiplicities is optimized to closely align with the data. At $\alpha=3$, the match significantly improves, providing a better representation of the observed data. Consequently, the cumulative distribution function (CDF) of the cluster size is computed for each event and each layer independently considering the position in the strip from Figure~\ref{fig:clustersize}(a) for both the X side and Y side in different layers. Consequently, the X-side cluster size and Y-side cluster size are generated using the acceptance-rejection method on the CDF utilizing the two correlated uniform random variables 'u' and 'v'. The modified simulated joint distribution is depicted in Figure~\ref{fig:joint_dist}(c). It is clear from Figure~\ref{fig:joint_dist}(b) that without Clayton copula, the correlation is very much different than data, in Figure~\ref{fig:joint_dist}(a). Although this procedure alters the correlation between X-side and Y-side multiplicities, it does not impact the marginal multiplicity distribution.

\begin{figure}[htbp]
\centering % \begin{center}/\end{center} takes some additional vertical space
	\begin{subfigure}[b]{0.3\textwidth}
        \centering
    	\includegraphics[width=\textwidth]{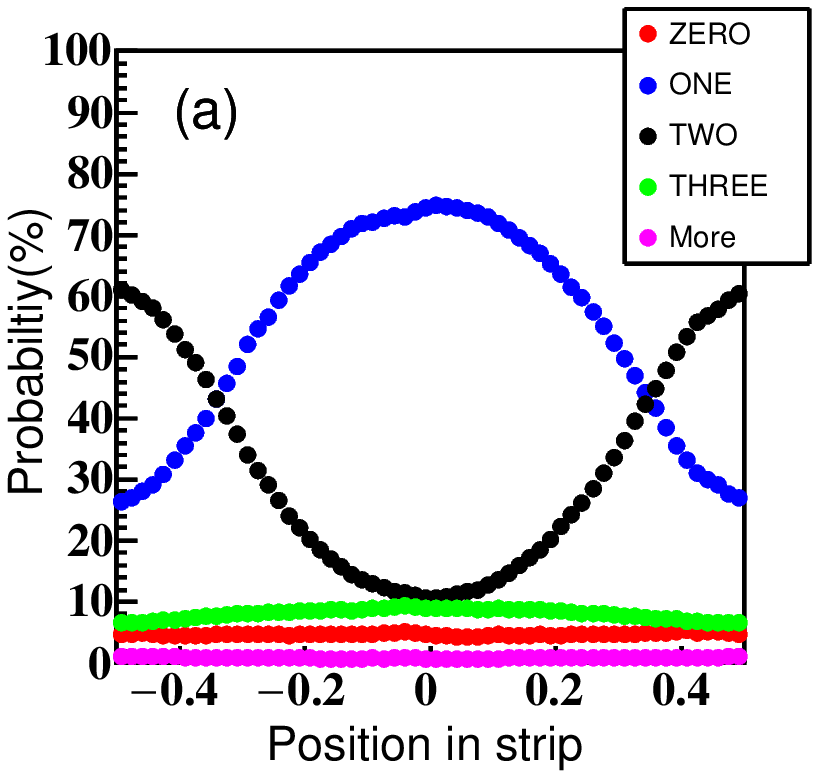}
    \end{subfigure}
\hfill
	\begin{subfigure}[b]{0.3\textwidth}
         \centering
         \includegraphics[width=\textwidth]{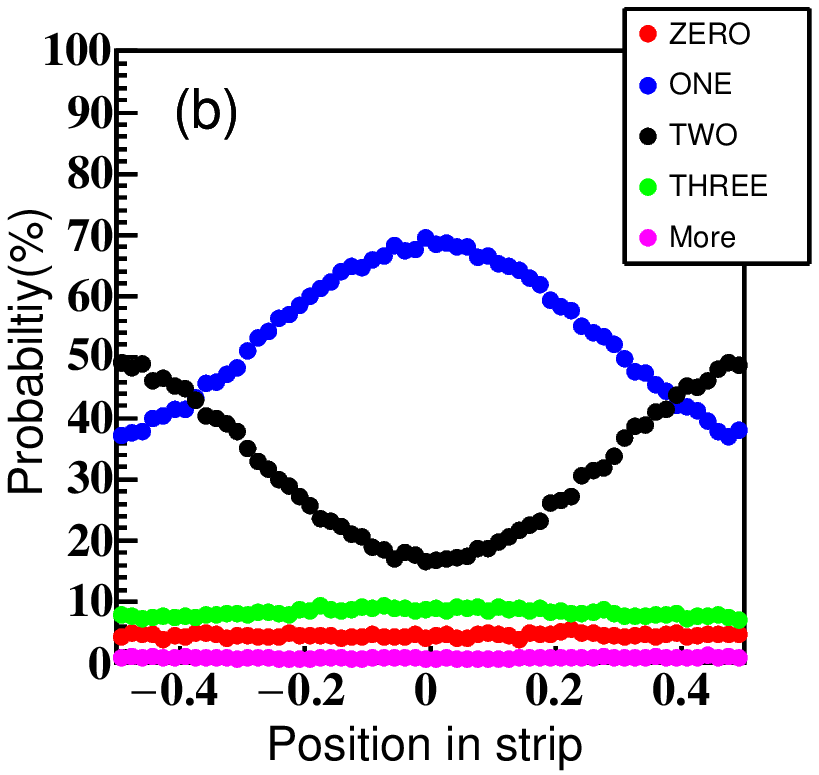}
    \end{subfigure}
\hfill
	\begin{subfigure}[b]{0.3\textwidth}
         \centering
         \includegraphics[width=\textwidth]{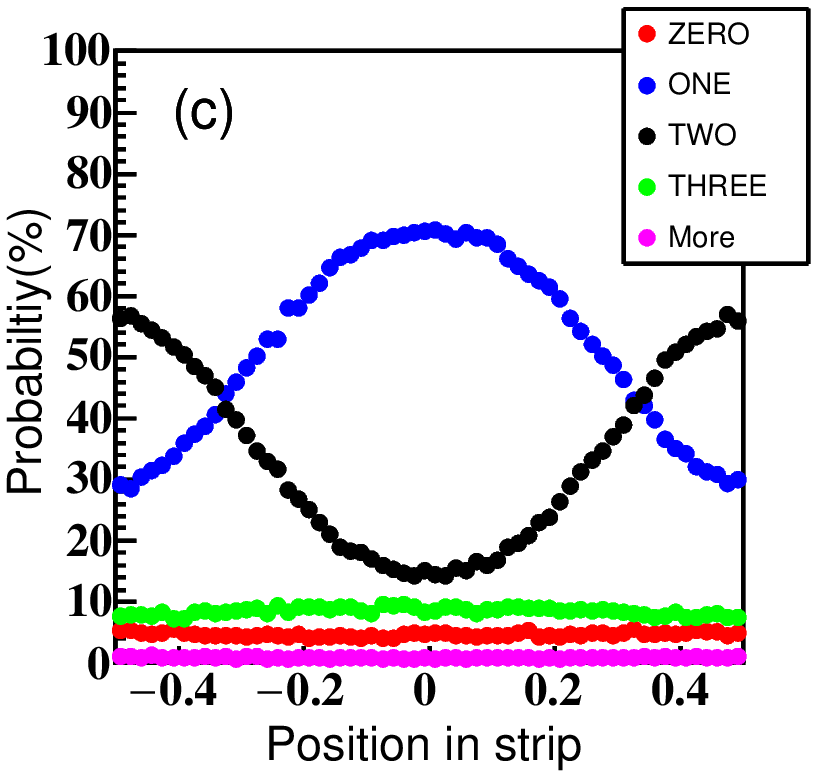}
    \end{subfigure}
\caption{\label{fig:deconvolution} The cluster size with respect to position in a strip (a) in data (b) in the simulation before deconvolution, (c) in the simulation after deconvolution.}
\end{figure}

In the process of determining the position within a strip, a straight-line fit is applied to the muon track, allowing for estimation of the muon position within a layer. However, this estimation is subject to fitting errors, resulting in the convolution of the cluster size distribution with respect to the position within the strip and the fitting error. The average position resolution is 7.2\,mm, while the fitting error ranges from 3.3\,mm in the outer layers to 1.8\,mm in the inner layers. To obtain accurate cluster size distributions, it is essential to perform a deconvolution process that effectively removes the contribution of the fitting error from the cluster size distribution. Prior to convolution, the distribution of cluster sizes in relation to position is represented in Figure~\ref{fig:deconvolution}(b) which does not match with data as shown in Figure~\ref{fig:deconvolution}(a). Following the application of the deconvolution process, the distribution undergoes modification, as illustrated in Figure~\ref{fig:deconvolution}(c). This deconvolution procedure effectively removes the influence of the fitting error, resulting in a refined representation of the cluster size distribution with respect to position.

In certain layers, certain strips have been masked due to large noise. Also, there are dead strips due to the failure of electronics. However, the absence of a signal in a particular strip does not imply that the avalanche process is not occurring within the RPC. When a muon passes through a region with dead strips, the RPC still generates signals in the neighboring strips. Simulating this phenomenon is a complex and intricate process. A relatively straightforward approach to recreate the exact conditions is to include the dead strips while calculating the multiplicity distribution as shown in Figure~\ref{fig:clustersize}(a).   But, those are excluded from the final list of digitised strip hits.

\section{Extracting and Reincorporating Noise Information}
\label{section:noise}
\begin{figure}[htbp]
\centering % \begin{center}/\end{center} takes some additional vertical space
\includegraphics[width=.6\textwidth]{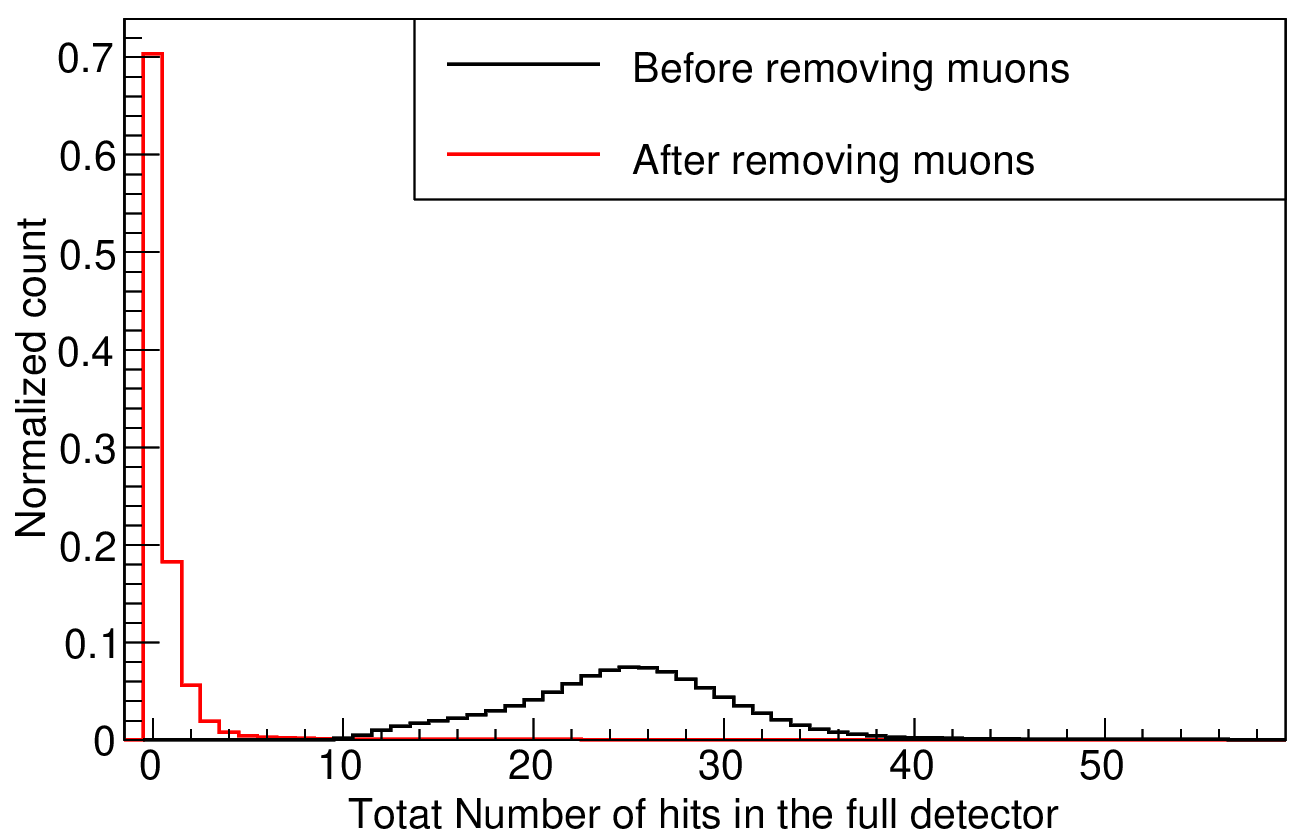}
\caption{\label{fig:totalnoise} The total noise hits distribution in the entire stack before and after removing muon hits.}
\end{figure}

\begin{figure}[htbp]
\centering % \begin{center}/\end{center} takes some additional vertical space
\includegraphics[width=.8\textwidth,origin=c]{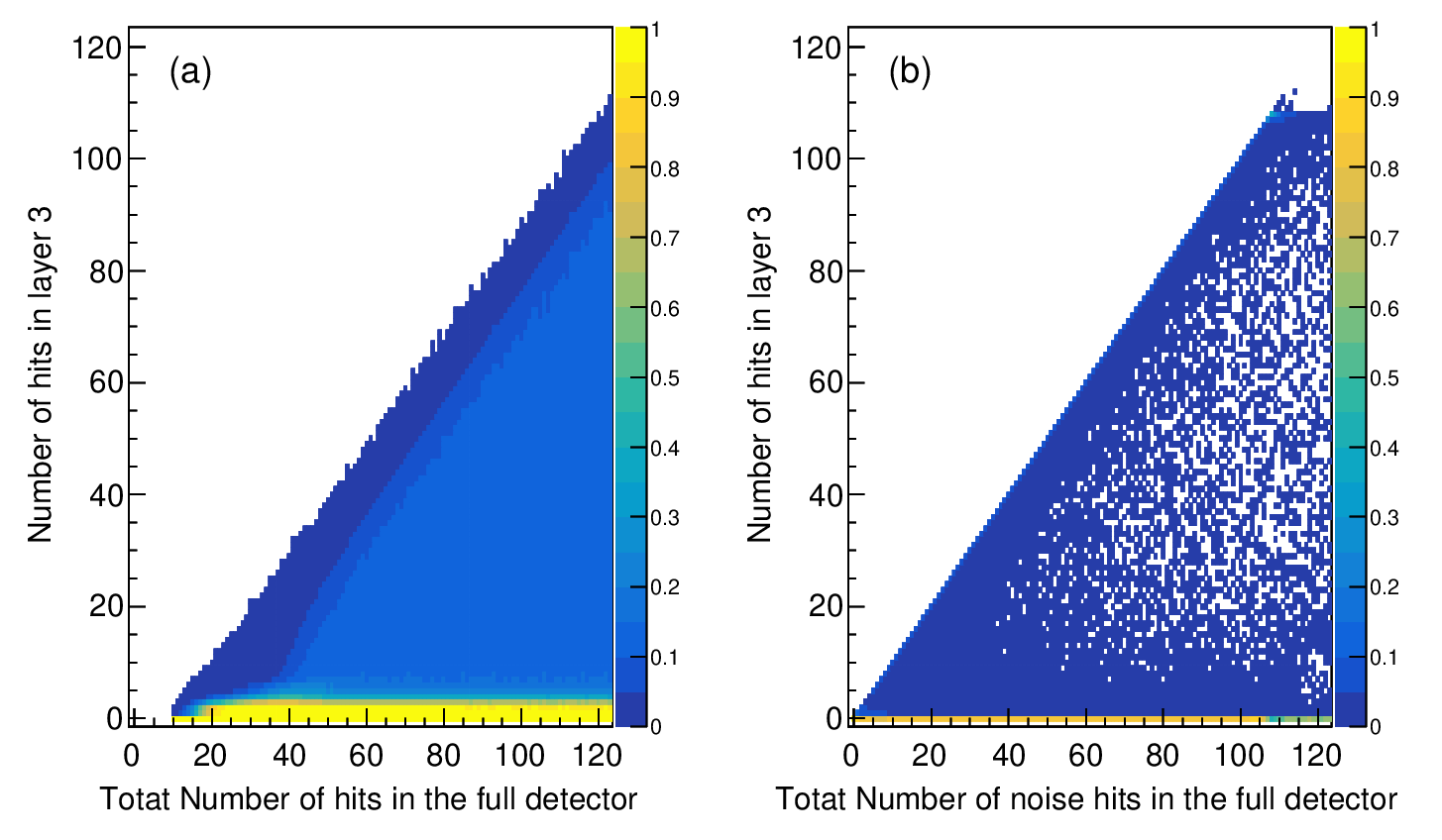}
% "\includegraphics" from the "graphicx" permits to crop (trim+clip)
% and rotate (angle) and image (and much more)
\caption{\label{fig:layernoise} The correlation between total number of hits in the entire stack
  to that in a specific layer, (a) before removing muon hits (b) after removing muon hits.}
\end{figure}

The noise information RPCs exhibit an average of approximately 80 hits per second per strip. During reconstruction, the data utilized is limited to a range of $\pm$10~ns relative to the average muon time, resulting in a noise occurrence of only 1 in 520 events for the entire stack comprising 1200 strips. However, the presence of cross-talk between channels leads to a higher noise fraction compared to random noise. Noise hits were identified as those hits that deviate from the muon hit by $\pm$7\,sigma or if the layer was not used in the fit, where sigma is the position residue without correcting for the fitting error. To analyze the distribution of noise hits %GMA fixed number to be replaced by extrapolation error 
across the entire stack, the total number of noise hits was extracted. Figure~\ref{fig:totalnoise} displays the distribution of the number of hits in the entire stack, both before and after removing the muon hits. Moreover, Figure~\ref{fig:layernoise} illustrates the correlation between the total number of noise hits and the number of noise hits in each layer.

  %  With remarkable precision, this study impeccably aligns the
  %various parameters of the detector, including residue measurements from each
  %layer, efficiencies, cluster sizes of strip multiplicities, misalignment and more, to
  %match the observation in data. A successful Monte Carlo(MC) simulation
  %of the cosmic muon in the miniICAL is presented here.

\begin{figure}[htbp]
\centering % \begin{center}/\end{center} takes some additional vertical space
\includegraphics[width=.6\textwidth,origin=c]{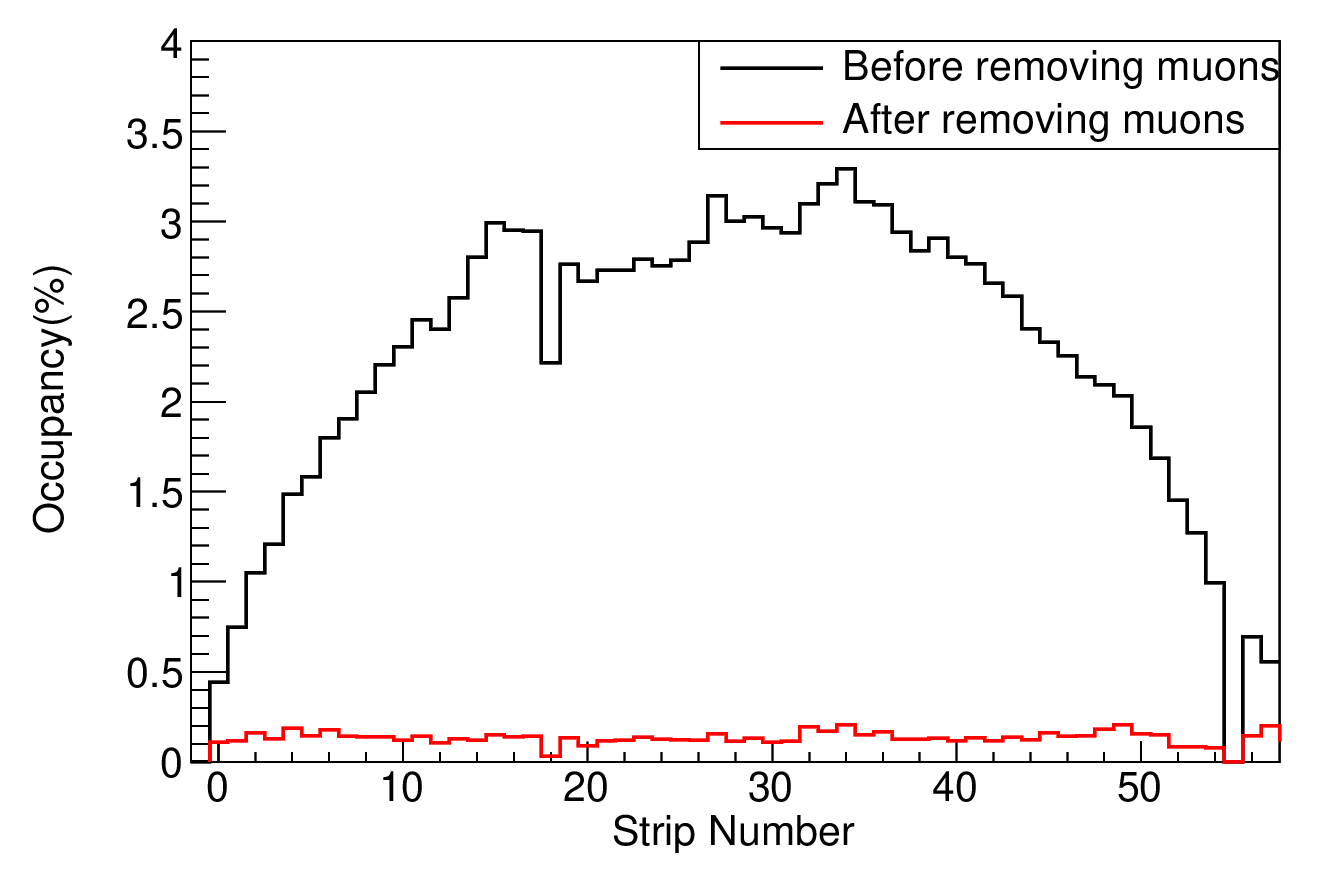}
% "\includegraphics" from the "graphicx" permits to crop (trim+clip)
% and rotate (angle) and image (and much more)
\caption{\label{fig:noiseoccupancy} The fraction of noise hits in layer 3 compared to the total hit occupancy.}
\end{figure}

To reintroduce noise into the simulation, the total number of noise hits for each event was stochastically sampled from the noise distribution as shown in Figure~\ref{fig:totalnoise}. Subsequently, a layer was randomly chosen, and the number of noise hits in that specific layer was determined from the histogram presented in Figure~\ref{fig:layernoise}(b). Next, the noise was randomly distributed based on the noise fraction associated with each strip. The process of selecting layers continues until the total number of noise hits matches the sampled value from the noise distribution, ensuring that the desired total noise hit count is achieved. The fraction of noise hits in a layer for each X-strip in contrast to the fraction of muon hits is shown in Figure~\ref{fig:noiseoccupancy}.

\begin{figure}[htbp]
\centering % \begin{center}/\end{center} takes some additional vertical space
\includegraphics[width=.6\textwidth,origin=c]{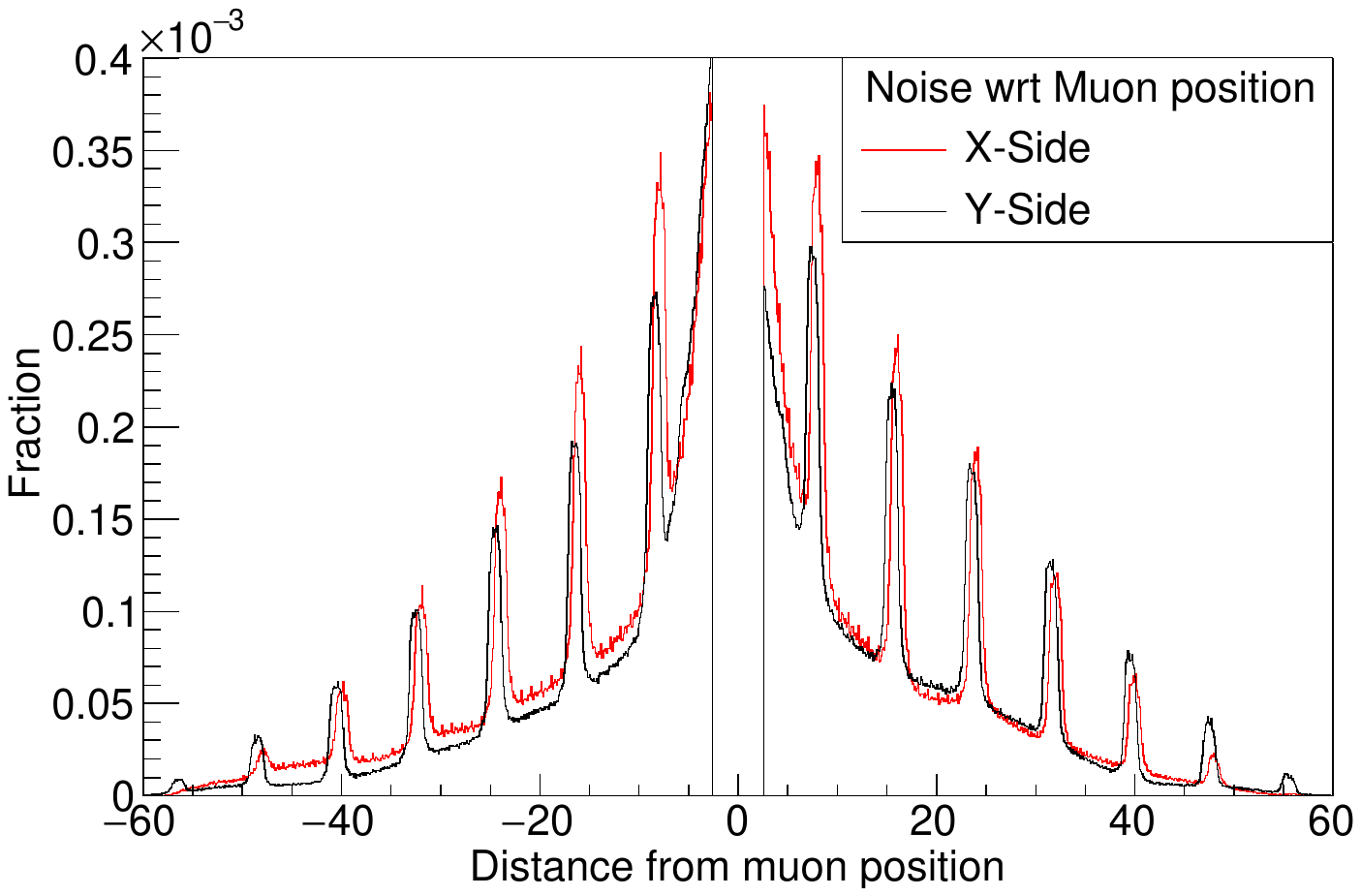}
% "\includegraphics" from the "graphicx" permits to crop (trim+clip)
% and rotate (angle) and image (and much more)
\caption{\label{fig:noise_eighth_strips} The distribution of distance to the noise hits in layer 3 from the muon extrapolated position.}
\end{figure}

In the process of time detection, the DAQ board employs an OR operation on specific, intermittent eighth strips on each side of the RPC. This is a necessary measure due to the limited number of channels in the Time to Digital Converter (TDC). However, this approach introduces cross-talk in each intermittent eighth channel. The distance to the noise hits from the muon extrapolated position after removing the muon hits is shown in Figure~\ref{fig:noise_eighth_strips}. If the muon passes through a layer in simulation, the noise is distributed based on this distribution in Figure~\ref{fig:noise_eighth_strips}. 

\begin{figure}[htbp]
\centering % \begin{center}/\end{center} takes some additional vertical space
\includegraphics[width=.8\textwidth,origin=c]{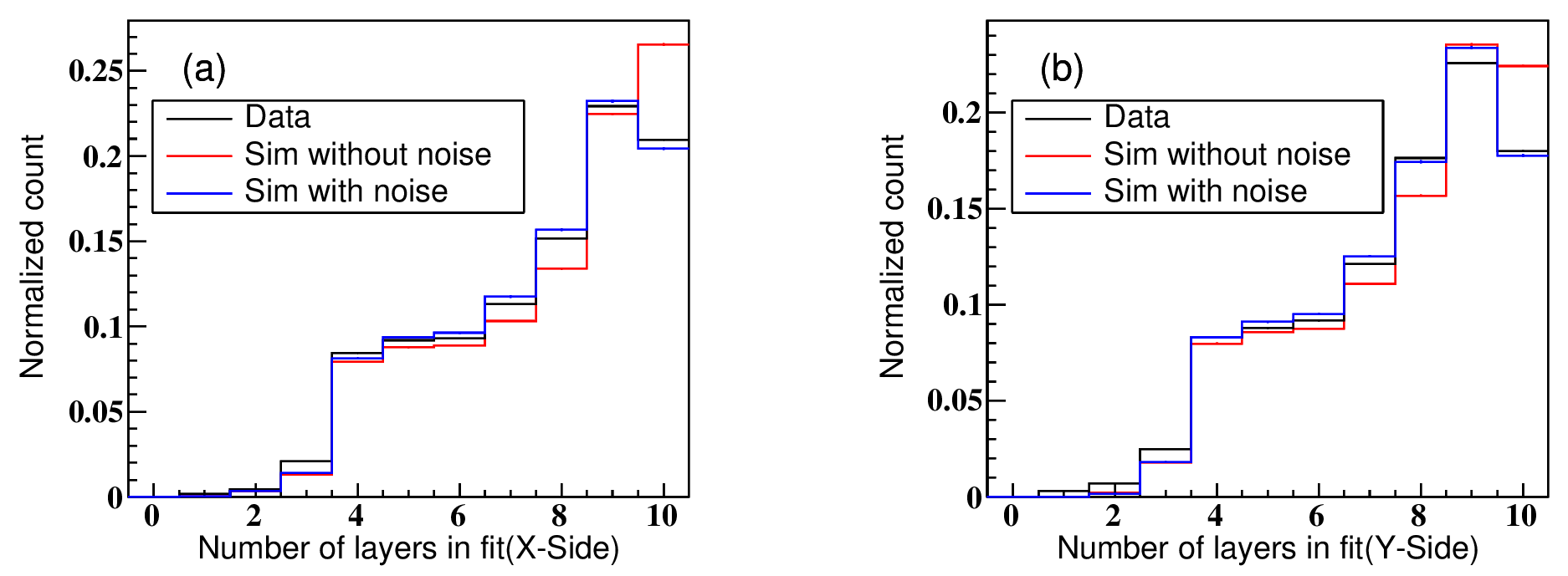}
% "\includegraphics" from the "graphicx" permits to crop (trim+clip)
% and rotate (angle) and image (and much more)
\caption{\label{fig:ndofold} The number of layers in the fit before and after introducing noise in the simulation, (a) X-side (b) Y-side.}
\end{figure}

The fidelity of noise in both data and simulation was evaluated through multiple criteria. First, the noise occupancy in the strips is compared between the data and Monte Carlo, ensuring their consistency. Additionally, the number of layers during fitting is compared between the data and MC, both before and after introducing noise in the MC. The results are illustrated in Figure~\ref{fig:ndofold}, providing a comprehensive analysis of the noise consistency between the two datasets.

The presence of noise does have a bias in the measurement of the zenith($\theta$) and azimuth($\phi$) distributions. Though an increase in noise reduces the precision of the measurement, the mean of these distributions and shapes within the uncertainty limit remain unaffected, maintaining their inherent characteristics and providing reliable information for further analysis.

%The data with magnetic field is fitted with a kalman fitter algorithm along with a pattern recognition algorithm for track finding. This process eliminates all the noise and the muon hit is used for fitting even if the layer contains a noise hit. Due to this there is not much difference in the case of ndf even if noise it there or not.

\begin{figure}[htbp]
\centering % \begin{center}/\end{center} takes some additional vertical space
\includegraphics[width=.75\textwidth,origin=c]{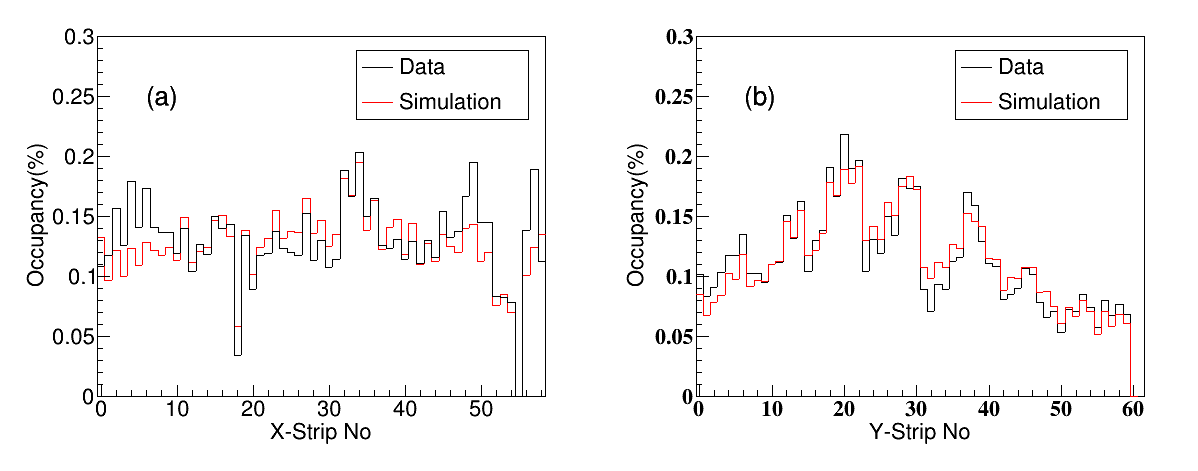}
% "\includegraphics[scale=•]{•}" from the "graphicx" permits to crop (trim+clip)
% and rotate (angle) and image (and much more)
\caption{\label{fig:nonmagnetic_noise} The comparison of noise occupancy in layer 3 of data when the magnetic field is OFF and the simulation with the noise taken from the data when the magnetic field is OFF, (a) X-side and (b) Y-side.}
\end{figure}

\begin{figure}[htbp]
\centering % \begin{center}/\end{center} takes some additional vertical space
\includegraphics[width=.75\textwidth,origin=c]{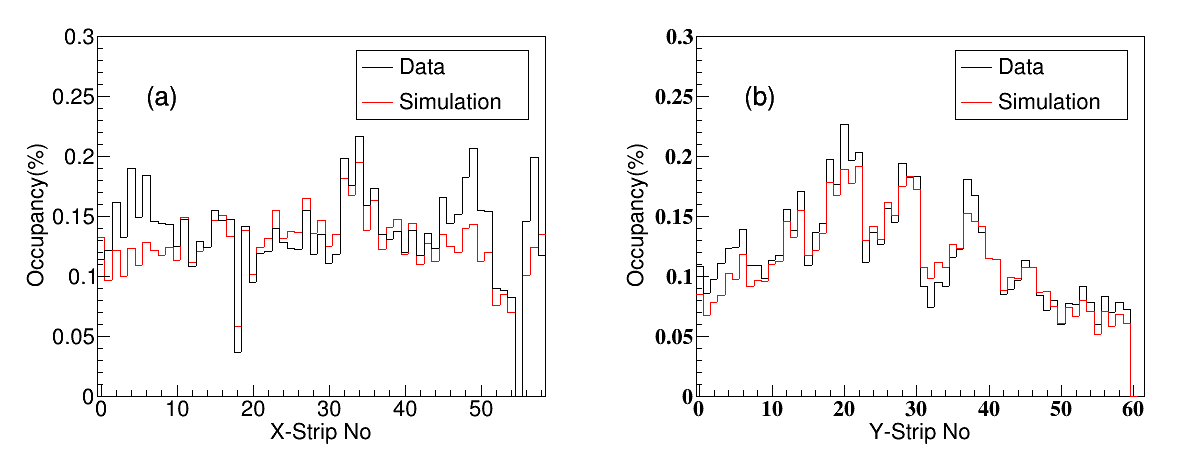}
% "\includegraphics[scale=•]{•}" from the "graphicx" permits to crop (trim+clip)
% and rotate (angle) and image (and much more)
\caption{\label{fig:magnetic_noise} The comparison of noise occupancy in layer 3 of data when the magnetic field is ON and the simulation with the noise taken from the data when the magnetic field is OFF, (a) X-side and (b) Y-side.}
\end{figure}

The level of noise occupancy in data is comparable between the presence and absence of the magnetic field. Figure~\ref{fig:nonmagnetic_noise} illustrates the comparison of noise occupancy in layer 3 between data and simulation, with the data acquired when the magnet is OFF. Figure~\ref{fig:magnetic_noise} illustrates the comparison of noise occupancy in layer 3 between data and simulation, with the data acquired under the influence of an active magnetic field.

\section{Incorporating Misalignment into Simulation}
\begin{figure}[htbp]
\centering % \begin{center}/\end{center} takes some additional vertical space
\includegraphics[width=.6\textwidth,origin=c]{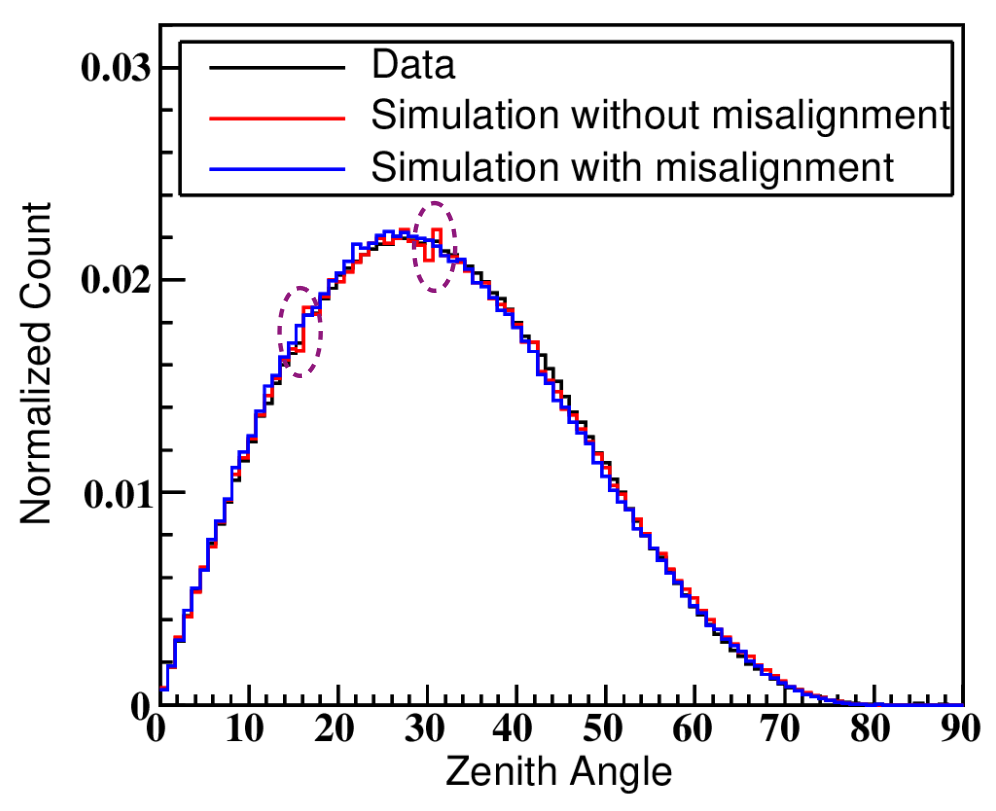}
% "\includegraphics" from the "graphicx" permits to crop (trim+clip)
% and rotate (angle) and image (and much more)
\caption{\label{fig:angle} The zenith angle distribution for data and MC without any misalignment in the MC and with misalignment applied in the MC.}
\end{figure}

The RPCs are positioned within the central region of iron layers. The alignment of the readout  panels of the RPCs was performed offline using the methodology outlined in the work by S. Pal et al.~\cite{sumanta_pal}. The time offset alignment procedure is detailed in the study conducted by~\cite{jim_paper1}. It is essential to accurately incorporate the misalignment in the simulation, as it notably impacts the azimuthal and zenith angle distributions. Ensuring the faithful reproduction of these distributions is crucial for an accurate comparison between the simulated and experimental data. An additional point worth noting is that the alignment procedure described above does not correct for shear in the detector. To achieve a faithful matching, it is necessary to re-implement the shifts in the Monte Carlo simulation. Figure~\ref{fig:angle} illustrates prominent peaks in the MC  compared to the data at specific angles. These peaks arise from the higher probability of cluster size 1 and when all layers exhibit a cluster size of 1, resulting in pronounced peaks.  The prominent peaks in the zenith angle distribution are observed at $tan^{-1}(3/10.1)$ and $tan^{-1}(6/10.1)$, where 3\,cm is the strip width and 10.1\,cm is the distance between layers. However, by incorporating the precise layer shifts in the simulation, the peaks are not observed in the distribution, as depicted in Figure~\ref{fig:angle}.

\section{Simulation with magnetic field ON}

\begin{figure}[htbp]
\centering % \begin{center}/\end{center} takes some additional vertical space
\includegraphics[width=.8\textwidth,origin=c]{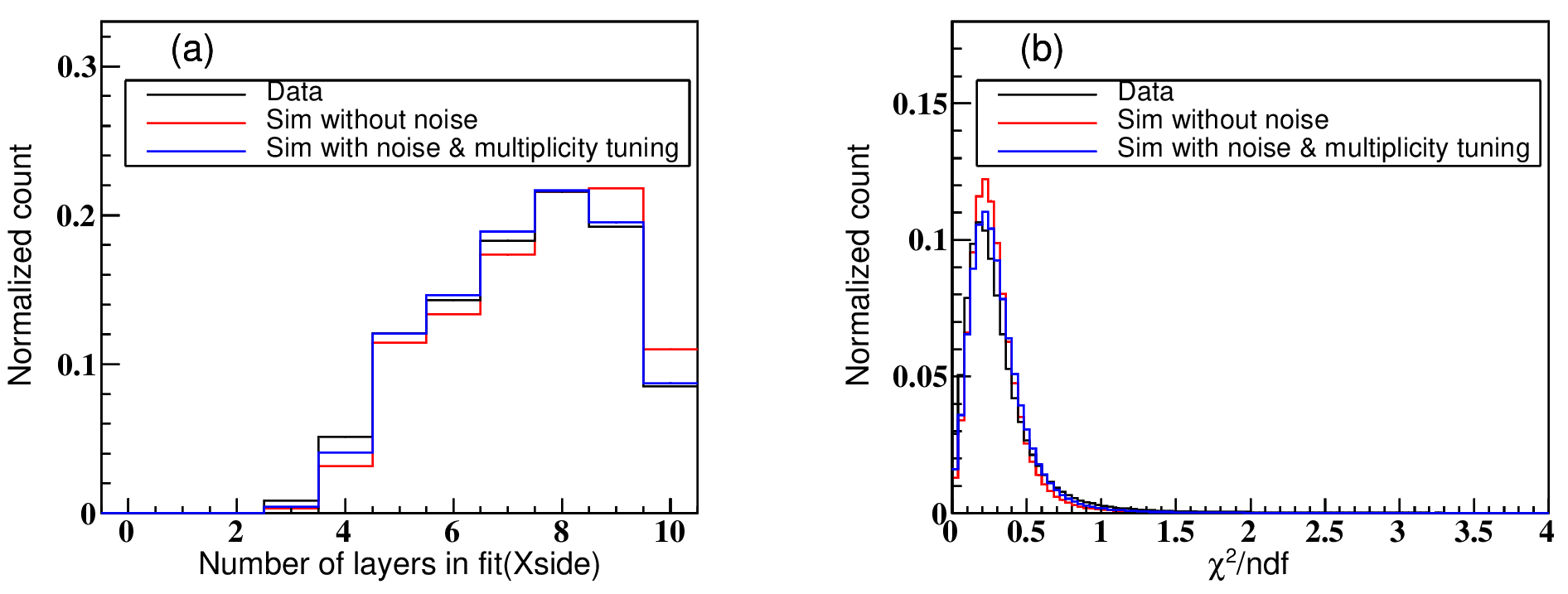}
% "\includegraphics" from the "graphicx" permits to crop (trim+clip)
% and rotate (angle) and image (and much more)
\caption{\label{fig:magneticNoise} (a) The number of layers in fit distribution and (b) the $\chi^2/ndf$ distribution for the case the magnetic field is ON, and for that reason the Kalman fitter algorithm is used for track fitting.}
\end{figure}

The magnetic field is administered in the Y direction, opposite in polarity, resulting in the XZ-plane being the bending plane. The track is fitted with a Kalman filter algorithm along with a pattern recognition algorithm for track finding~\cite{kolahal}. Figure~\ref{fig:magneticNoise} illustrates the comparison of the number of layers in fit and $\chi^2/ndf$ between the data and simulation. This comparison is presented both prior to the inclusion of noise and position-based multiplicity, as well as after their incorporation.

\section{Comparison of data and MC}
This simulation framework is developed by incorporating those detector features one by one. As it was mentioned in the introduction, we are unaware that any other experiments used only RPC for the momentum measurement from the curvature and did not find details for RPC noise simulation.
We have done simulations with one or more features missing to show the importance of their inclusion in the simulation. Figure~\ref{fig:ndofold} compares characteristics of events in data and MC without any magnetic field and Figure~\ref{fig:magneticNoise} shows the same with magnetic field. Remarkably, with the inclusion of all these features, a very good match was observed between data and MC, while using inputs from the data without magnetic field to simulate MC with magnetic field.

\section{Conclusion}
A detailed framework of RPCs detector simulation is developed which includes the properties of the RPCs. These properties are inefficiencies, strip multiplicities due to the trajectory of the charged particle as a function of local strip coordinates as well as global coordinates, correlation of efficiency on both sides of the readout chain, correlated and uncorrelated noise rates, effect of dead strips, etc. After incorporating all these in the real detector simulation the benchmark distributions in the data were well reproduced in the simulation framework. The correlation between X-Side and Y-Side multiplicity is also incorporated. Even in the same system, the characteristics of any two RPCs are different and all those are simulated independently, though the framework is common. It is possible that there are features, that are not observed here but may appear in the future, and the simulation framework would need to be updated. But this simulation framework is established for the mini-ICAL experiment and now it can be used to extract physics with the mini-ICAL detector, particularly, where one needs to unfold the observed distributions to true one, e.g., charge-dependent muon momentum.

%\appendix
%\section{Appendix}
%Please always give a title also for appendices.

\acknowledgments

We extend our appreciation to the INO collaboration for their indispensable support. We sincerely thank the individuals at IICHEP, Madurai TIFR, Mumbai, and BARC, Mumbai, whose dedicated contributions were integral to the construction and operation of mini-ICAL. Special thanks to Prof. V.M. Datar, who has also given valuable suggestions to improve the quality of this draft.



\begin{thebibliography}{99}

\bibitem{ichep2018} Gobinda Majumder {\it et. al.}, \emph{Design, construction and performance of magnetised mini-ICAL detector module}, \href{https://inspirehep.net/literature/1748958} { PoS ICHEP2018 (2019) 360}

\bibitem{raveendrababu} 
K. Raveendrababu {\it et. al.}, \emph{Effect of electrical properties of glass electrodes on the performance of {RPC} detectors for the {INO}-{ICAL} experiment}, \href{https://doi.org/10.1088/1748-0221/11/08/p08024} {Journal of Instrumentation {\bf 11} (2016) P08024}.

\bibitem{jim_paper1}
J.M. John {\it et. al.}, \emph{Improving time and position resolutions of RPC detectors using time over threshold information}, \href{https://doi.org/10.1088/1748-0221/17/04/P04020}{Journal of Instrumentation {\bf 17} (2022) P04020}.

\bibitem{achrekar}
Achrekar S. {\it et. al.}, \emph{Electronics, Trigger and Data Acquisition Systems for the INO ICAL Experiment. In: Liu ZA. (eds) Proceedings of International Conference on Technology and Instrumentation in Particle Physics 2017. TIPP 2017}, \href{https://doi.org/10.1007/978-981-13-1313-4_55}{Springer Proc. Phys. {\bf 212} (2018) 291-295}.

\bibitem{nino} F. Anghinolfi {\it et. al.}, \emph{NINO: an ultra-fast and low-power front-end amplifier/discriminator ASIC designed for the multigap resistive plate chamber}, \href{https://doi.org/10.1016/j.nima.2004.07.024}{Nucl. Instr. Meth. Phys. Res. {\bf A533} (2004) 183-187}.

\bibitem{mandar} M. N. Saraf {\it et. al.}, \emph{Electronics and DAQ for the Magnetized mini-ICAL Detector at IICHEP}, \href{https://doi.org/10.1007/978-981-33-4408-2_108} {Springer Proc.Phys. {\bf 261} (2021) 779-786}.

\bibitem{surya_multiplicity} 
Suryanarayan Mondal {\it et. al.}, \emph{Study of particle multiplicity of cosmic ray events using 2\,m$\times$2\,m resistive plate chamber stack at IICHEP-Madurai}, \href{https://doi.org/10.1007/s10686-020-09685-6}{Experimental Astronomy {\bf 51} (2021) 17-32}. 


\bibitem{Pethuraj_2017}
S. Pethuraj {\it et. al.}, \emph{Measurement of cosmic muon angular distribution and vertical integrated flux by 2\,m $\times$ 2\,m RPC stack at IICHEP-Madurai}, \href{https://dx.doi.org/10.1088/1475-7516/2017/09/021}{Journal of Cosmology and Astroparticle Physics {\bf 09} (2017) 021}.

\bibitem{corsika}
Heck, D. {\it et. al.}, \emph{CORSIKA: A Monte Carlo code to simulate extensive air showers}, 
\href{https://inspirehep.net/literature/469835}{FZKA-6019 (1998)}.

\bibitem{geant4}
Agostinelli, S. {\it et. al.}, \emph{GEANT4--a simulation toolkit}, \href{https://doi.org/10.1016/S0168-9002(03)01368-8}{Nucl. Instrum. Meth. A{\bf 506} (2003) 250-303}.

\bibitem{sumanta_pal}
S. Pal {\it et. al.}, \emph{Study of the directionality of cosmic muons using the INO-ICAL prototype detector}, \href{https://doi.org/10.1016/j.nima.2013.09.025} {Nuclear Instruments and Methods in Physics Research Section A{\bf 753} (2014) 88-93}.

\bibitem{kolahal} 
Kolahal Bhattacharya{\it et. al.}, \emph{Error propagation of the track model and track fitting strategy for the Iron CALorimeter detector in India-based neutrino observatory},  \href{https://doi.org/10.1016/j.cpc.2014.09.003} {Computer Physics Communications {\bf 185} (2014) 3259-3268}.

%\bibitem{b}
%Author, \emph{Title}, \emph{J. Abbrev.} {\bf vol} (year) pg.

%\bibitem{c}
%Author, \emph{Title},
%arxiv:1234.5678.

%\bibitem{d}
%Author, \emph{Title},
%Publisher (year).




% Please avoid comments such as "For a review'', "For some examples",
% "and references therein" or move them in the text. In general,
% please leave only references in the bibliography and move all
% accessory text in footnotes.

% Also, please have only one work for each \bibitem.


\end{thebibliography}
\end{document}